\documentclass[twocolumn,showpacs,preprintnumbers,amsmath,amssymb]{revtex4}

\usepackage{dcolumn}
\usepackage{bm}
\usepackage{amsmath,amssymb,amsfonts,latexsym,fancyhdr,graphicx}

\newcommand{\ket}[1]{\displaystyle{|#1\rangle}}
\newcommand{\bra}[1]{\displaystyle{\langle#1|}}
\newcommand{\al}{\alpha}
\newcommand{\Om}{\Omega}

\newcommand{\G}{\Gamma}
\newcommand{\g}{\gamma}
\newcommand{\si}{\sigma}
\newcommand{\la}{\lambda}

\begin{document}
\title{Reservoir cross-over in entanglement dynamics}
\author{L. Mazzola,$^1$ S. Maniscalco,$^1$ K.-A. Suominen,$^1$ and B. M. Garraway$^2$}
\affiliation{$^1$Department of Physics and Astronomy, University of
Turku, FI-20014 Turun yliopisto, Finland\\$^2$Department of Physics
and Astronomy, University of Sussex, Falmer, Brighton, BN1 9QH,
United Kingdom}

\date{\today}

\begin{abstract}
We study the effects of spontaneous emission on the entanglement
dynamics of two qubits interacting with a common Lorentzian
structured reservoir. We assume that the qubits are initially prepared in
a Bell-like state. We focus on the strong coupling regime and study
the entanglement dynamics for different regions of the spontaneous
emission decay parameter. This investigation allows us to explore
the cross-over between common and independent reservoirs in
entanglement dynamics.
\end{abstract}

\pacs{03.67.Bg; 03.65.Ud; 03.65.Yz; 42.50.-p}

\maketitle

\section{Introduction}

Since its discovery by Yu and Eberly in 2004, considerable interest
has been devoted to the phenomenon of early-stage disentanglement,
also called entanglement sudden death (ESD) \cite{Yu}. Indeed not
only is this phenomenon important from a fundamental point of view,
but it is likely to play a role in many applications of quantum
information theory and technology \cite{Nielsen,Harbook}. For
example, ESD might represent a threat to building a quantum
computer. So far ESD has been predicted in many different
theoretical systems such as pairs of qubits
\cite{Bellomo,Ficek,Etrapping,PRAESD}, continuous variable system
\cite{Paz}, subset of multiple qubits and spin chains. It has also
been observed experimentally in two different contexts
\cite{Almeida,Laurat}.

In general, a key factor determining the dynamics is the type of
environment in which the system of interest is immersed. With just
a pair of qubits, differences appear in the entanglement
evolution, depending on the properties of the environment. For
example, when a bipartite system interacts with one reservoir (i.e.\
the two parts interact with the same reservoir), correlations
between the two parts are created because of the reservoir-mediated
interaction. These correlations cannot arise if each of the parts talks to only
its own environment, and if there is no direct coupling present between them.
On the other hand, if the correlation time of the bath is long, the
environment keeps track of the dynamics of the small system, and
revivals of entanglement may appear. These are the so-called memory
effects, typical of non-Markovian dynamics, appearing not only in
common, but also in independent reservoirs.

In Ref. \cite{PRAESD} we have studied the exact dynamics of two
qubits interacting with a common Lorentzian-structured, non-Markovian
reservoir. Here we generalize these results and take into account the
spontaneous emission from the two qubits. This model describes, for example,
the situation of two atoms strongly coupled to the same
one-dimensional high-Q cavity, in which both the atoms can also
independently emit a photon in any direction outside the cavity. The
uni-dimensional high-Q cavity constitutes a common non-Markovian
reservoir for the qubits, while the independent spontaneous emission
of a photon in a flat continuum of modes can be seen as a consequence
of the interaction with two independent Markovian reservoirs.

The introduction of spontaneous emission for the two atoms allows us
to study a rather realistic situation, characterized by different
regimes of entanglement dynamics.
In particular, we want to explore the cross-over between
common and independent reservoir dynamics.

In Ref. \cite{Kari} the generation of entanglement for two trapped ions
coupled to a high-finesse cavity has been studied. In that case, not only
the cavity losses, but also the entire atomic level structure (the
ions are coupled to the cavity mode via a Raman scheme in a
$\Lambda$-configuration) is taken into account. The aim of that
work is to study realistic experimental conditions under which the
collective Dicke model can be implemented in an ion-cavity QED context,
so only states with one excitation in the atomic system are
considered. Here we study entanglement dynamics when the atomic
system is prepared in Bell-like states with one and two excitations.

In the following sections we first introduce the master
equation describing the system of interest (section \ref{sec:model}), then we study the
entanglement dynamics for initial Bell-like states for different
regions of parameters (section \ref{sec:ent_dynamics}), and finally we interpret our results and make
some conclusive remarks in section \ref{sec:conclusion}.

\section{The model}
\label{sec:model}

\begin{figure}
\begin{center}
\includegraphics[width=4.3cm]{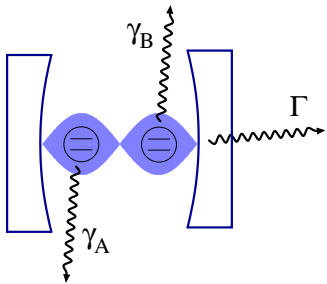}
\caption{\label{fig:setup}(Color online)
Diagrammatic representation of the model we consider.
Two atoms couple to a cavity mode which leaks to the environment at
a rate $\Gamma$. In addition, each atom can independently
radiate directly to the environment at the rates $\gamma_A$,
$\gamma_B$.
}
\end{center}
\end{figure}

\begin{figure}[!]
\begin{center}
\includegraphics[width=8.6cm]{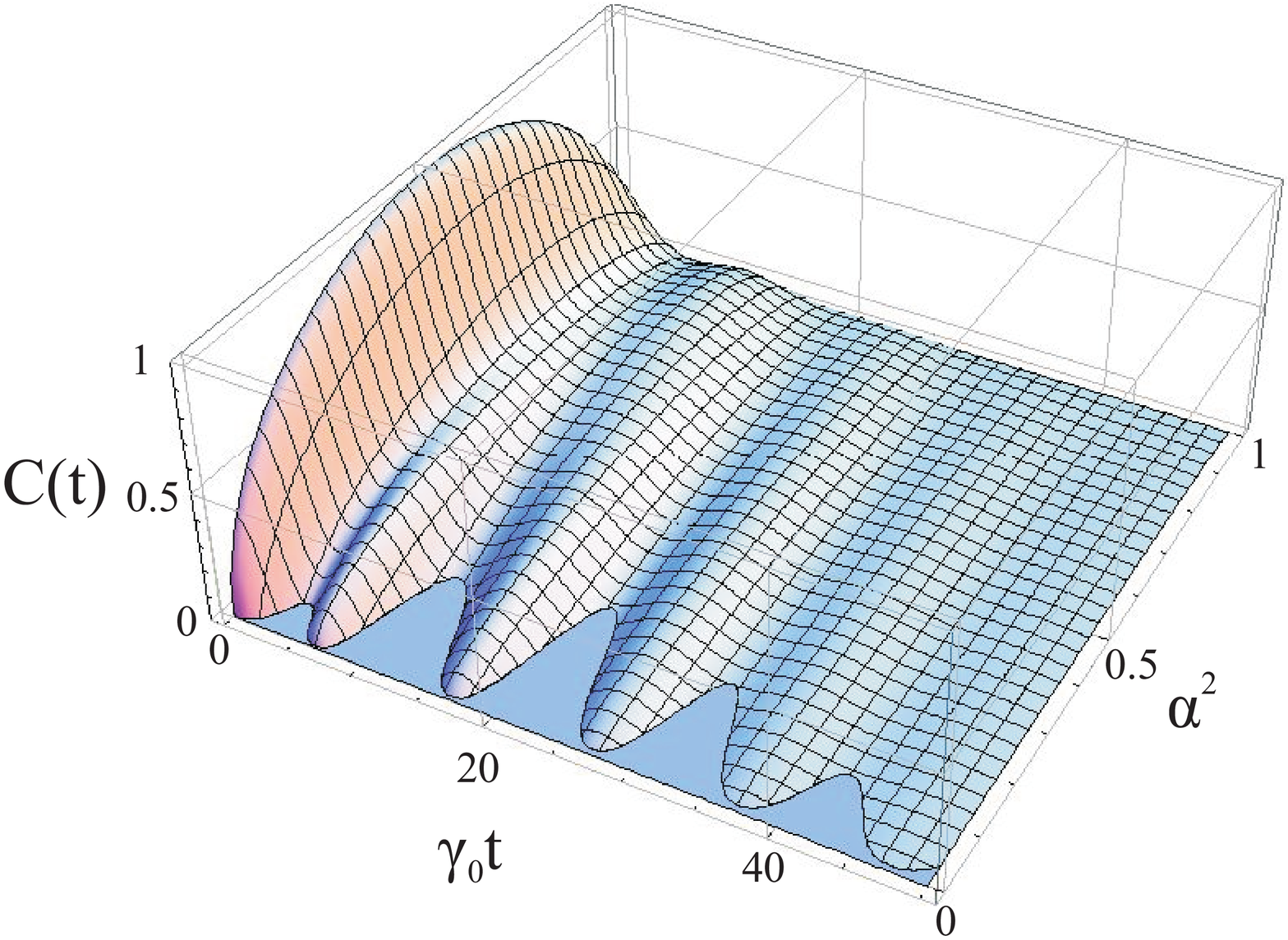}
\caption{\label{fig:PRAESD}(Color online) Concurrence as a function
of scaled time and $\al^2$ for two atoms prepared in the state
\eqref{Psi} and in the absence of spontaneous emission.
(Taken from Ref.\ \cite{PRAESD}.)}
\end{center}
\end{figure}

We consider two two-level systems (qubits) interacting with the same
leaky cavity in rotating-wave approximation, and emitting
independently outside the cavity due to spontaneous emission
(see figure \ref{fig:setup}). The
qubits have the same transition frequency, and they are equally and
resonantly coupled with the leaky cavity. In Ref. \cite{PRAESD} we
have studied the exact dynamics of two qubits interacting with the
same Lorentzian-structured reservoir, i.e.\ a lossy resonator. In that
work we found that, for a certain class of initial states, the
entanglement dynamics exhibits regions of sudden death and
resurrections. We have interpreted these results as a consequence of
the memory effects of the non-Markovian environment and of the
reservoir-mediated interaction between the qubits. For the sake of
clarity, and as a reference point, we report in figure \ref{fig:PRAESD}
the basic result for the concurrence. We solved
the dynamics using the pseudomode approach \cite{Barry97,Breuer}.
The pseudomode master equation describes a coherent Tavis-Cummings
type interaction between the qubits and the pseudomode, and the
dissipative part involves the pseudomode only which leaks into a
Markovian reservoir. For a high-Q cavity the pseudomode is basically
identified as the cavity mode.

Here we generalize such a master equation by adding phenomenologically
the dissipative spontaneous emission part for the two qubits.
Contrary to Ref. \cite{PRAESD} we do not use the representation in terms
of super-radiant and sub-radiant states, but we keep the natural
basis $\{\ket{00},\ket{10},\ket{01},\ket{11}\}$ for the two qubits,
as the super-radiant basis has no advantage when the independent
spontaneous emission processes break the symmetry of the system. Thus the
master equation of the system in the interaction picture is
\begin{eqnarray}\label{ME}
\frac{\partial\tilde{\rho}}{\partial t}=-\imath[H,\tilde{\rho}]
-\frac{\G}{2}(a^{\dag}a\tilde{\rho}+\tilde{\rho}a^{\dag}a-2a\tilde{\rho}a^{\dag})\nonumber\\
-\frac{\g_{A}}{2}(\si_{+}^{A}\si_{-}^{A}\tilde{\rho}+\tilde{\rho}\si_{+}^{A}\si_{-}^{A}-2\si_{-}^{A}\tilde{\rho}\si_{+}^{A})\nonumber\\
-\frac{\g_{B}}{2}(\si_{+}^{B}\si_{-}^{B}\tilde{\rho}+\tilde{\rho}\si_{+}^{B}\si_{-}^{B}-2\si_{-}^{B}\tilde{\rho}\si_{+}^{B}),
\end{eqnarray}
where $\tilde{\rho}$ is the density matrix of the qubits plus the
cavity mode and
\begin{equation}
H=\Omega[(\si_{+}^{A}+\si_{+}^{B})a+(\si_{-}^{A}+\si_{-}^{B})a^{\dag}].
\end{equation}
Here, $\si_{\pm}^{A}$ and $\si_{\pm}^{B}$ are, respectively, the
Pauli raising and lowering operators for the atoms A and B, $a$ and
$a^{\dagger}$ are the annihilation and creation operators for the
cavity mode, $\Om$ is the cavity-qubits coupling constant, $\G$ is
the cavity decay rate, and $\g_{A/B}$ are the spontaneous emission rates
for atoms A and B. For simplicity we consider $\g_{A}=\g_{B}=\g_{S}$
and the coupling $\Omega$ to be the same for both qubits.

We assume that the environment is at zero temperature, so there are
at most two excitations in the total system. The basis we use to
write the density matrix elements of $\tilde{\rho}$ is
$\{\ket{000},\ket{001},\ket{002},\ket{100},\ket{101},\ket{010},\ket{011},\ket{110}\}$,
where the first and second digit indicate ground (0) and excited (1)
state of qubit A and B respectively, and the third one indicates the
number of excitations inside the cavity. We solve numerically the 64
differential equations for the density matrix elements, and then, to find
the dynamics of the two-qubit system only, we trace out the cavity
mode degree of freedom to find the reduced density matrix $\rho$.

Once we have the reduced qubits density matrix we can derive
the entanglement dynamics. To quantify entanglement we use the Wootters
concurrence~\cite{Wootters}, defined as
$C(t)=\textrm{max}\{0,\sqrt{\la_{1}}-\sqrt{\la_{2}}-\sqrt{\la_{3}}-\sqrt{\la_{4}}\}$,
where $\{\la_{i}\}$ are the eigenvalues of the matrix
$R=\rho(\si_{y}^{A}
\otimes\si_{y}^{B})\rho^{\ast}(\si_{y}^{A}\otimes\si_{y}^{B})$, with
$\rho^{\ast}$ denoting the complex conjugate of $\rho$ and
$\si_{y}^{A/B}$ are the Pauli matrices for atoms $A$ and $B$. This
quantity attains its maximum value of 1 for maximally entangled
states and vanishes for separable states.

We assume the qubits are prepared in Bell-like states. The evolution then
drives such initial pure states into a mixed state having an \lq\lq
X\rq\rq\ form
\begin{equation}\label{rhot}
   \rho(t)=\left(
      \begin{array}{cccc}
        a(t) & 0 & 0 & w(t) \\
        0 & b(t) & z(t) & 0 \\
        0 & z^{*}(t) & c(t) & 0 \\
        w^{*}(t) & 0 & 0 & d(t) \\
      \end{array}
   \right).
\end{equation}
For this class of states the concurrence assumes a simple analytic
expression
\begin{equation}\label{conc}
   C(t)=\mathrm{max}\{0,C_{1}(t),C_{2}(t)\},
\end{equation}
where
\begin{eqnarray}\label{C1C2}
   C_{1}(t)&=2|w(t)|-2\sqrt{b(t) c(t)},\\
   C_{2}(t)&=2|z(t)|-2\sqrt{a(t) d(t)},
\end{eqnarray}
where coherences give a positive contribution to $C_{1}(t)$ and
$C_{2}(t)$ and hence to the concurrence, while the negative parts involve
populations only.
Of course, any initial state in the ``X'' form of Eq.~\eqref{rhot},
pure, or mixed, would lead to the same expressions for the
concurrence. For states not in this form we would have to perform a
numerical evaluation of the concurrence.

\section{Entanglement dynamics}
\label{sec:ent_dynamics}

We investigate entanglement dynamics for the following initial Bell-like states of
the qubits:
\begin{equation}\label{Phi}
   \ket{\Phi}=\al\ket{10}+e^{i\theta}(1-\al^2)^{1/2}\ket{01},
\end{equation}
and
\begin{equation}\label{Psi}
   \ket{\Psi}=\al\ket{00}+e^{i\theta}(1-\al^2)^{1/2}\ket{11}.
\end{equation}
With appropriate choices of $\theta$ and $\alpha$ these states
include the usual four Bell states.

Our aim is to understand the interplay between common non-Markovian
reservoir and independent Markovian reservoirs in entanglement
dynamics. To do so we consider a strong coupling between the
qubits and the cavity, and then we change the spontaneous emission decay
rate $\gamma_S$.

A Bell-like state with \emph{one} excitation as in Eq.\ \eqref{Phi} never presents
entanglement sudden death \cite{PRAESD}. In the strong coupling
regime and in absence of spontaneous emission the entanglement
exhibits oscillations which are damped in time. The introduction of
spontaneous emission from the two qubits leads to an additional damping
in the oscillations. The result is that the larger the spontaneous
emission rate is, the smaller the entanglement revivals are; for
$\gamma_{S}\gtrsim 3\Omega$ the oscillations are completely killed and
the entanglement does not revive anymore.

The dynamics of the Bell-like state involving \emph{two} excitations, Eq.\ \eqref{Psi}, is
more interesting. First of all, let us recall the two limiting
cases: no coupling with the cavity and no spontaneous emission. We
obtain the Markovian independent reservoirs dynamics when setting
the coupling between the qubits and the cavity to zero. In this case
there are two different regions of the entanglement dynamics: for
$\al^2<1/2$ entanglement dies suddenly, and for $\al^2\geq1/2$
entanglement decays exponentially. In the second limiting case, i.e.\ with
cavity coupling and in absence of spontaneous emission, the qubits
interact exclusively with a common non-Markovian reservoir. In this
case entanglement presents a much richer dynamics with oscillations
for every value of $\al^2$ and a series of dark periods (ESD
regions) and resurrections of entanglement for $\al^2\lesssim1/4$, as
shown in figure \ref{fig:PRAESD} \cite{PRAESD}.

Here we explore entanglement dynamics for different values of the
spontaneous emission rate $\g_{S}$, fixing the strong coupling
between the qubits and the cavity. In particular we set the cavity
parameters to $\Om=0.2\g_{0}$ and $\G=\sqrt{0.05}\g_{0}$ where
$\g_{0}$ is the decay parameter in the case of an infinitely broad
cavity. These parameters correspond to experimentally feasible
conditions in the context of, for example, trapped ions \cite{Blatt}
or circuit-QED \cite{Sillanpaa}. The results do not depend on the
relative phase $\theta$.

\begin{figure}[!]
\begin{center}
\includegraphics[width=8.6cm]{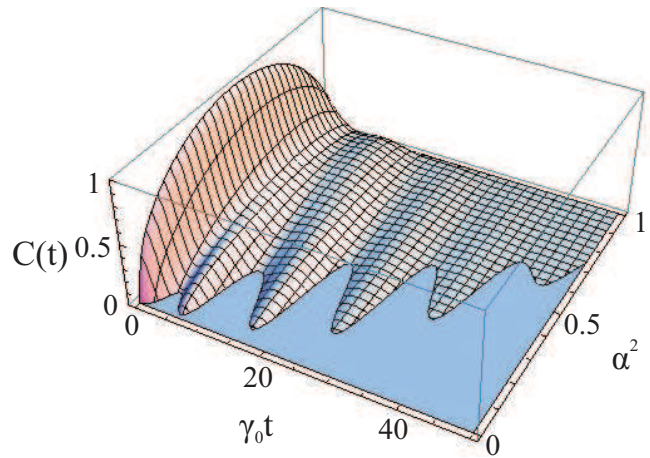}
\caption{\label{fig:gdiv10}(Color online) Concurrence as a function
of scaled time and $\al^2$ for two atoms prepared in state
\eqref{Psi}. Here the spontaneous emission parameter $\g_{S}$ is ten
times smaller than the cavity-qubits coupling constant $\Om$.}
\end{center}
\end{figure}

In figure \ref{fig:gdiv10} we present the entanglement dynamics for two
qubits prepared in the state $\eqref{Psi}$ with a weaker spontaneous emission rate
$\g_{S}=\Om/10$. We notice that for short times the dynamics is not
affected by spontaneous emission (compare  with figure
\ref{fig:PRAESD}). As time passes the effects of spontaneous emission
become prominent: i.e.\ the oscillations get smaller in amplitude, and the
ESD region dramatically increases until, for a time around
$t\thickapprox2/\gamma_{S}\thickapprox100\g_{0}$, entanglement is
dead irrespective of the value of $\al^2$.

A clearer explanation of the dynamics can be found by looking at the
evolution of the density matrix elements. For a two-photon Bell-like
state the concurrence is given by $C_{1}(t)$, where the two-photon coherence
gives a positive contribution, and the populations of the state with
one excitation in one of the qubits give a negative contribution. Looking at
the differential equations for the density
matrix elements $\tilde{\rho}$, Eq. \eqref{ME}, we notice that the terms describing
spontaneous emission cause a faster decay of two-photon coherence.
On the other hand the population of the states $\ket{10}$ and
$\ket{01}$ is less affected, since spontaneous emission not only
adds a decay channel from $\ket{10}$/$\ket{01}$ to $\ket{00}$ but
also one from $\ket{11}$ to $\ket{10}$/$\ket{01}$. As a consequence,
at a certain time depending on the parameter $\al^2$, the two-photon
coherence will be smaller than the one-excitation populations and
entanglement will be definitely lost.

\begin{figure}[!]
\begin{center}
\includegraphics[width=8.6cm]{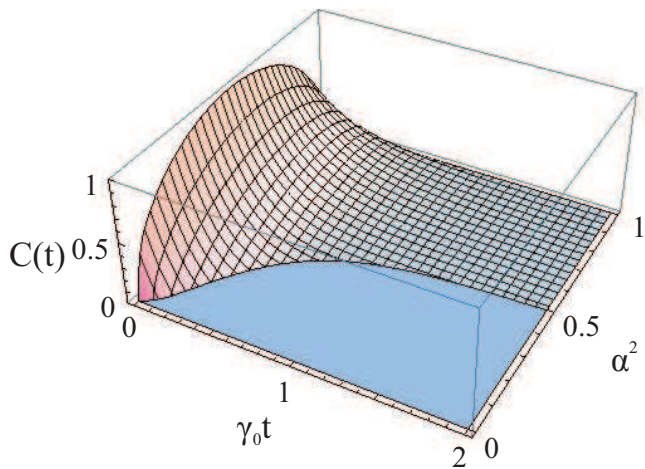}
\caption{\label{fig:gper10}(Color online) Concurrence as a function
of scaled time and $\al^2$ for two atoms prepared in state
\eqref{Psi}. Here the spontaneous emission parameter $\g_{S}$ is ten
times larger than the cavity-qubits coupling constant $\Om$.}
\end{center}
\end{figure}

In the case of a strong spontaneous emission rate,
$\g_{S}=10\Om$, the entanglement dynamics tends to be the same as the
independent Markovian reservoirs case, as figure \ref{fig:gper10}
shows. Even though the coupling with the cavity is strong,
the entanglement does not revive and the spontaneous emission damps down
any entanglement oscillations completely. We also see that the entanglement vanishes
exponentially for $\al^2\geq1/2$ and dies suddenly for $\al^2<1/2$.

\begin{figure}[!]
\begin{center}
\includegraphics[width=8.6cm]{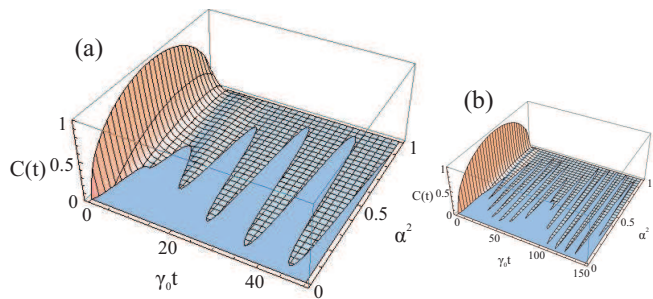}
\caption{\label{fig:gequal}(Color online) Concurrence as a function
of scaled time and $\al^2$ for two atoms prepared in state
\eqref{Psi} for small (a) and long (b) time-scales. Here the
spontaneous emission parameter $\g_{S}$ is equal to the
cavity-qubits coupling constant $\Om$.}
\end{center}
\end{figure}

More delicate is the case of a comparable cavity-qubits coupling and
spontaneous emission decay rate. Figure \ref{fig:gequal} shows
the entanglement dynamics at different time-scales for $\g_{S}=\Om$.
Compared to the case with no spontaneous emission, the dynamics is quite
different from the very beginning. Indeed the region of ESD appears to be
strongly increased. However, later on the coherent interaction
with the cavity makes entanglement revive and, as time passes,
revivals appear for a wider and wider range of the $\al^2$ parameter,
as shown in figure \ref{fig:gequal} (a). For longer times (figure
\ref{fig:gequal} (b)), the range of $\al^2$ over which we find
multiple entanglement revivals shrinks, and then expands again
and, in fact, we find that this pattern repeats many times.
Clearly,
as time passes the amplitude of the oscillations and revivals
dramatically decreases, for example, for $t\sim150/\g_{0}$ the
concurrence is of the order of $10^{-6}$. The details of the
dynamics change a lot depending on the particular value of $\g_{S}$.
However, for $\g_{S}$ of the order of $\Om$, some general features of
the dynamics can be identified. For example, for $\al^2\rightarrow0$
and $\al^2\rightarrow1$ the dynamics of entanglement is mainly
controlled by spontaneous emission, resembling the independent
Markovian reservoirs case. For intermediate values of
$\al^2$ a number of death and revival periods follow one after the other
with decreasing intensity. This is a sign of the non-Markovian
backaction of the cavity and of the cavity-mediated coupling between
the qubits as well. In general the long-time picture looks like the
superimposition of death and revivals on the independent Markovian
reservoir dynamical pattern.

\section{Conclusion}
\label{sec:conclusion}

We have studied the entanglement dynamics of two qubits interacting
with a leaky cavity (shared non-Markovian reservoir) and emitting
independently via spontaneous emission (independent Markovian
reservoirs). Starting with the exact master equation of Ref.
\cite{PRAESD} describing the non-Markovian dynamics with the cavity,
we have phenomenologically added two dissipative terms for the
qubits. Depending on the ratio between the spontaneous emission
parameter $\g_{S}$ and cavity-qubit coupling constant $\Om$,
different regimes in entanglement dynamics can be identified. When
the spontaneous emission rate is small, the entanglement dynamics is
not very much affected for short times, but, as time passes, the
region of sudden death spreads out until for long times the
entanglement is lost for every value of $\al^2$. When $\g_{S}$ is
much larger than $\Om$, the interaction with the cavity does not
play any role. The two dissipative spontaneous emission terms in Eq.
\eqref{ME} control the dynamics, preventing any possibility of
revivals in entanglement. The intermediate region when $\g_{S}$ and
$\Om$ are of the same order of magnitude shows a more complex
behaviour exhibiting dynamical elements of both the two limiting
cases.

We emphasize that the method presented here can be applied
to any general initial state of the qubits, and not only to
Bell-like states. In particular, it is interesting to see the
effects of mixedness of the initial state on the
entanglement dynamics given that the dynamics depends
dramatically on the state of preparation of the qubits.
Here we just mention the case of an extended Werner-like
state with two excitations of the form
$\rho=r\ket{\Psi}\bra{\Psi}+(1-r)\mathbb{I}/4$. This state
is characterized by an entangled part, formed from the
Bell-like state of Eq.  \eqref{Psi}, and a maximally mixed
part, identified with the unit matrix. Clearly, due to the
mixedness of the state, the initial amount of entanglement is
smaller than for the state $\ket{\Psi}$ alone. The presence
of the mixed part causes a faster death of entanglement.
However, for small values of the $\g_{S}$ parameter the
entanglement exhibits oscillations and, as $\g_{S}$
increases, features of the reservoir cross-over appear.

In realistic experimental ion-cavity QED conditions it is not always
possible to limit the losses of the system to cavity-losses, and
often spontaneous emission needs to be taken into account. In our
work we have demonstrated how the phenomenon of spontaneous emission
comes into play in the entanglement dynamics of strongly interacting
cavity QED systems.

\acknowledgments

The authors thank W. Lange and J. Piilo for useful discussions. This
work has been financially supported by M. Ehrnrooth Foundation,
V\"{a}is\"{a}l\"{a} Foundation, Turku University Foundation, Turku
Collegium of Science and Medicine, and the Academy of Finland
(projects 108699, 115682, 115982).


\begin{thebibliography}{25}

\bibitem{Yu}
T. Yu and J. H. Eberly, Phys. Rev. Lett. \textbf{93}, 140404 (2004),
T. Yu and J. H. Eberly, Science \textbf{323}, 598 (2009).

\bibitem{Nielsen}
M. A. Nielsen and I. L. Chuang, \emph{Quantum Computation and
Quantum Information}, (Cambridge University Press, Cambridge,
England, 2000); S. Stenholm and K.\,-\,A. Suominen, Quantum Approach
to Informatics, (John Wiley \& Sons, NJ, 2005).

\bibitem{Harbook}
S. Haroche and J.-M. Raimond, \emph{Exploring the Quantum: Atoms,
Cavities, and Photons}, (OUP, Oxford, 2006).

\bibitem{Bellomo}
B. Bellomo \textit{et al.}, Phys. Rev. Lett. \textbf{99}, 160502
(2007).

\bibitem{Ficek}
Z. Ficek and R. Tana\'{s}, Phys. Rev. A \textbf{74}, 024304 (2006).

\bibitem{Etrapping}
B. Bellomo \textit{et al.}, Phys. Rev. A \textbf{78}, 060302(R)
(2008).

\bibitem{PRAESD}
L. Mazzola, S. Maniscalco, J. Piilo, K.-A. Suominen and B. M.
Garraway, Phys. Rev. A \textbf{79}, 042302 (2009); L. Mazzola, S.
Maniscalco, J. Piilo and K.-A. Suominen, arXiv:0904.2857.

\bibitem{Paz}
J. P. Paz and A. J. Roncaglia, Phys. Rev. Lett. \textbf{100}, 220401
(2008).

\bibitem{Almeida}
M. P. Almeida \textit{et al.}, Science \textbf{316}, 579 (2007).

\bibitem{Laurat}
J. Laurat \textit{et al.}, Phys. Rev. Lett. \textbf{99}, 180504
(2007).

\bibitem{Kari}
K. Harkonen, F. Plastina, and S. Maniscalco, arXiv:0907.0778.

\bibitem{Barry97}
B. M. Garraway, Phys. Rev. A \textbf{55}, 2290 (1997).

\bibitem{Breuer}
H.-P. Breuer and F. Petruccione, \emph{The Theory of Open Quantum
Systems} (OUP, Oxford, 2002).

\bibitem{Wootters}
W. K. Wootters, Phys. Rev. Lett. \textbf{80}, 2245 (1998).

\bibitem{Blatt}
H. H\"{a}ffner \textit{et al.}, Nature \textbf{431}, 643 (2005); R.
Blatt and D. Wineland, Nature \textbf{453}, 1008 (2008); J. Majer
\textit{et al.}, Nature \textbf{449}, 443 (2005); G. R.
Guth\"{o}hrlein \textit{et al.}, Nature \textbf{414}, 49 (2001); A.
Wallraff \textit{et al.}, Nature \textbf{431}, 162 (2004); D. L.
Moehring \textit{et al.}, Nature \textbf{449}, 68 (2007).

\bibitem{Sillanpaa}
M. A. Sillanp\"{a}\"{a} \textit{et al.}, Nature \textbf{449}, 438
(2007).

\end{thebibliography}
\end{document}